\documentclass[twocolumn,prl,aps,showpacs]{revtex4-1} \usepackage[dvips]{graphicx}
\usepackage{longtable}
\usepackage{amsmath}

\usepackage{color}
\begin{document}
\title{Fluctuation diagnostics of the electron self-energy: Origin of the pseudogap physics}                                  

\author{O.~Gunnarsson,$^1$ T.~Sch\"afer,$^2$ J.~P.~F.~LeBlanc,$^{3,4}$ E.~Gull,$^4$ 
 J.~Merino,$^5$
G.~Sangiovanni,$^6$ G.~Rohringer,$^2$ and A.~Toschi$^2$ } 
\affiliation{
$^1$ Max-Planck-Institut f\"ur Festk\"orperforschung, Heisenbergstra{\ss}e 1, D-70569 Stuttgart, Germany \\ 
$^2$Institute of Solid State Physics, Vienna University of Technology, A-1040 Vienna, Austria \\
$^3$ Max-Planck-Institute for the Physics of Complex Systems, D-01187 Dresden, Germany\\ 
$^4$ Department of Physics, University of Michigan, Ann Arbor, Michigan 48109, USA \\
$^5$ Departamento de F\'isica Te\'orica de la Materia Condensada, IFIMAC Universidad Aut\'onoma de Madrid, Madrid 28049, Spain \\
$^6$ Institute of Physics and Astrophysics, University of W\"urzburg, D-97070 W\"urzburg, Germany}
 
\begin{abstract}
We demonstrate how to identify which physical processes dominate the low-energy spectral functions of correlated electron systems. We obtain an unambiguous classification through an analysis of the equation of motion for the electron self-energy in its charge, spin and particle-particle representations. Our procedure is then employed to clarify the controversial physics responsible for the appearance of the pseudogap in correlated systems. We illustrate our method by examining the attractive and repulsive Hubbard model in two-dimensions. In the latter, spin fluctuations are identified as the origin of the pseudogap, and we also explain why $d-$wave pairing fluctuations play a marginal role in suppressing the low-energy spectral weight, independent of their actual strength.

 \end{abstract}
\date{\today} 
%PACS: 71.10.-w: Theories and models of many-electron systems

%71.27.+a: Strongly correlated systems

%71.10.Fd: Lattice fermion models. Hubbard models
\maketitle 

{\sl Introduction.} -- Correlated electron systems display some of the most fascinating phenomena in condensed matter physics, but their understanding still represents a formidable challenge for theory and experiments. For photoemission \cite{ARPESrev} or STM \cite{STMrev,STSrev} spectra, which measure single-particle quantities, information about correlation is encoded in the electronic self-energy $\Sigma$. However, due to the intrinsically many-body nature of the problems, even an exact knowledge of $\Sigma$ is {\sl not} sufficient for an unambiguous identification of the underlying physics. A perfect example of this is the pseudogap observed in the single-particle spectral functions of underdoped cuprates \cite{Timusk1999}, and, more recently, of their nickelate analogues \cite{Uchida2011}. Although relying on different assumptions, many theoretical approaches provide self-energy results compatible with the experimental spectra. This explains the lack of a consensus about the physical origin of the pseudogap: In the case of cuprates, the  pseudogap has been attributed to spin-fluctuations \cite{spinflu1, spinflu2, spinflu3, spinflu4,spinflu5}, preformed pairs \cite{preform1,preform2,preform3,preform4, preform5}, Mottness \cite{Stanescu2003, Imada2013}, and, recently, to the interplay with charge fluctuations \cite{SilvaNeto2014,YangRice,RiceYang,Comin} or to Fermi-liquid scenarios \cite{Mirzaei2013}. The existence and the role of ($d-$wave) superconducting fluctuations \cite{preform1, preform2,preform3,preform4, preform5} in the pseudogap regime are still openly debated  for the basic model of correlated electrons, the Hubbard model. 
%Recently, a dynamical cluster approximation (DCA) \cite{DCA} study found \cite{Gull2012} only a weak response to an external $d-$wave pairing field in the pseudogap phase, while other DCA calculations evidenced \cite{pseudogap,pseudogap1} strong short-range $d$-wave pair fluctuations.

Experimentally, the clarification of many-body physics is augmented by  a simultaneous investigation at the two-particle level, i.e., via neutron scattering \cite{INSrev}, infrared/optical spectroscopy \cite{OPTrev}, muon-spin relaxation \cite{MUrev}, etc. Analogously, theoretical studies of $\Sigma$ can also be supplemented by a corresponding analysis at the two-particle level. In this paper, we study the influence of the two-particle fluctuations on $\Sigma$ via its equation of motion. We apply this method of ``fluctuation diagnostics'' to identify the role played by different collective modes in the pseudogap physics.   

{\sl Self-energy decomposition.} -- We emphasize that all concepts and equations below are applicable within any theoretical approach in which the self-energy and the two-particle vertex are calculated without {\sl a priori} assumptions of a predominant type of fluctuations. This includes, e.g., quantum Monte-Carlo (QMC) methods such as lattice QMC \cite{bss}, functional renormalization group \cite{fRGrev}, parquet approximation \cite{parquet,parquet1,Yang2009}, and cluster extensions \cite{Maier} of the dynamical mean field theory (DMFT) \cite{DMFT, DMFTrev} such as the cellular-DMFT \cite{CDMFTlichtenstein,CDMFT} or the dynamical cluster approximation (DCA) \cite{DCA}. Within diagrammatic extensions \cite{DGA,DGA1,DF,DF1,nanoDGA,multiscale,1PI} of DMFT, our analysis is applicable if parquet-like diagrams are included \cite{DFparquet,DMF2RG,nanoparquet}.   

The self-energy describes all scattering effects of {\sl one} added/removed electron, when propagating through the lattice. In correlated electronic systems, these scattering events originate from the Coulomb interaction among the electrons themselves, rather than from the presence of an external potential. Therefore, $\Sigma$ is entirely determined by the full {\sl two-particle} scattering amplitude $F$. The formal relation between $F$ and $\Sigma$ is known as Dyson-Schwinger equation of motion (EOM) \cite{Abrikosov}. In the important case of a purely local interaction (as in the Hubbard model \cite{Hubbard,Hubbard1,Hubbard2}), this reads (in the paramagnetic phase)
\begin{equation}
\label{eq:1}
\Sigma(k)=  \frac{Un}{2}   
-\frac{U}{\beta^2 N}\sum_{k',q} \, F_{\uparrow \downarrow}(k,k',q) \, g(k')g(k'+q)g(k+q),  
\end{equation}
where $U$ is the (bare) Hubbard interaction, $n$ the electronic density, $g$ the electron 
Green's function, $\beta =1/T$ the inverse temperature, 
and $N$ the normalization of the momentum summation (we adopt the notation 
$k=(\nu,{\bf K})$/$q=(\omega, {\bf Q})$ for the fermionic/bosonic Matsubara frequencies $\nu$/$\omega$ and momenta ${\bf K}$/${\bf Q}$, see the supplementary material for details). 
Finally, $F_{\uparrow \downarrow}$ is the full scattering amplitude (vertex) between electrons with opposite spins: It consists of repeated two-particle scattering events in {\sl all} possible configurations compatible with energy/momentum/spin conservation. Therefore it contains the complete information of the {\sl two}-particle correlations of the system. Yet, much of the information encoded in $F_{\uparrow \downarrow}$ about the {\sl specific} physical processes determining $\Sigma$ is washed out by averaging over all two-particle scattering events, i.e., by the summations on the r.h.s. of Eq.~(\ref{eq:1}). Hence, an unambiguous identification of the physical role played by the underlying scattering/fluctuation processes requires a ``disentanglement'' of the EOM. The most obvious approach would be a direct decomposition of the full scattering amplitude $F_{\uparrow \downarrow}$ of Eq.~(\ref{eq:1}) in {\sl all} possible fluctuation channels, the so-called parquet \cite{parquet,parquet1,parquet2,Yang2009} decomposition, where different contributions to  $F$ are identified in terms of their two-particle reducibility. Inserting this in Eq.~(\ref{eq:1}), the contributions to $\Sigma$ can be attributed to the different channels, allowing for a clear physical understanding. We find, however, that this approach only works in the weakly correlated regime (small $U$, large doping, high $T$): For stronger correlations, the numerical decomposition procedure becomes highly unstable, due to divergences in the two-particle {\sl irreducible} vertex functions, recently discovered in the Hubbard and Falicov-Kimball models \cite{divergence, unpublished, Janis}, see also \cite{Kozik2014}. For example, in our DCA calculations for the two-dimensional ($2D$) Hubbard model the breakdown of the parquet decomposition of $\Sigma$ occurs at {\sl lower} values of $U$ (or larger values of doping) than those for which pseudogap physics is numerically observed.   

In this paper we present an alternative route that can be followed to circumvent this problem. Our idea exploits the freedom of employing formally equivalent analytical representations of the EOM. For instance, by means of SU(2) symmetry  and ``crossing relations'' (see, e.g., \cite{RVT,Tam2013}), we can  express $F_{\uparrow \downarrow}$ in Eq.~(\ref{eq:1}) in terms of the corresponding vertex functions of the spin/magnetic $F_{sp}=F_{\uparrow \uparrow} - F_{\uparrow \downarrow}$ and charge/density  $F_{ch}= F_{\uparrow \uparrow} +F_{\uparrow \downarrow}$ sectors. Analogously, a rewriting in terms of the particle-particle sector notation is  done via $F_{pp}(k,k',q)=F_{\uparrow \downarrow}(k,k',q-k-k')$.  
\begin{figure}
{\rotatebox{-90}{\resizebox{9.0cm}{!}{\includegraphics {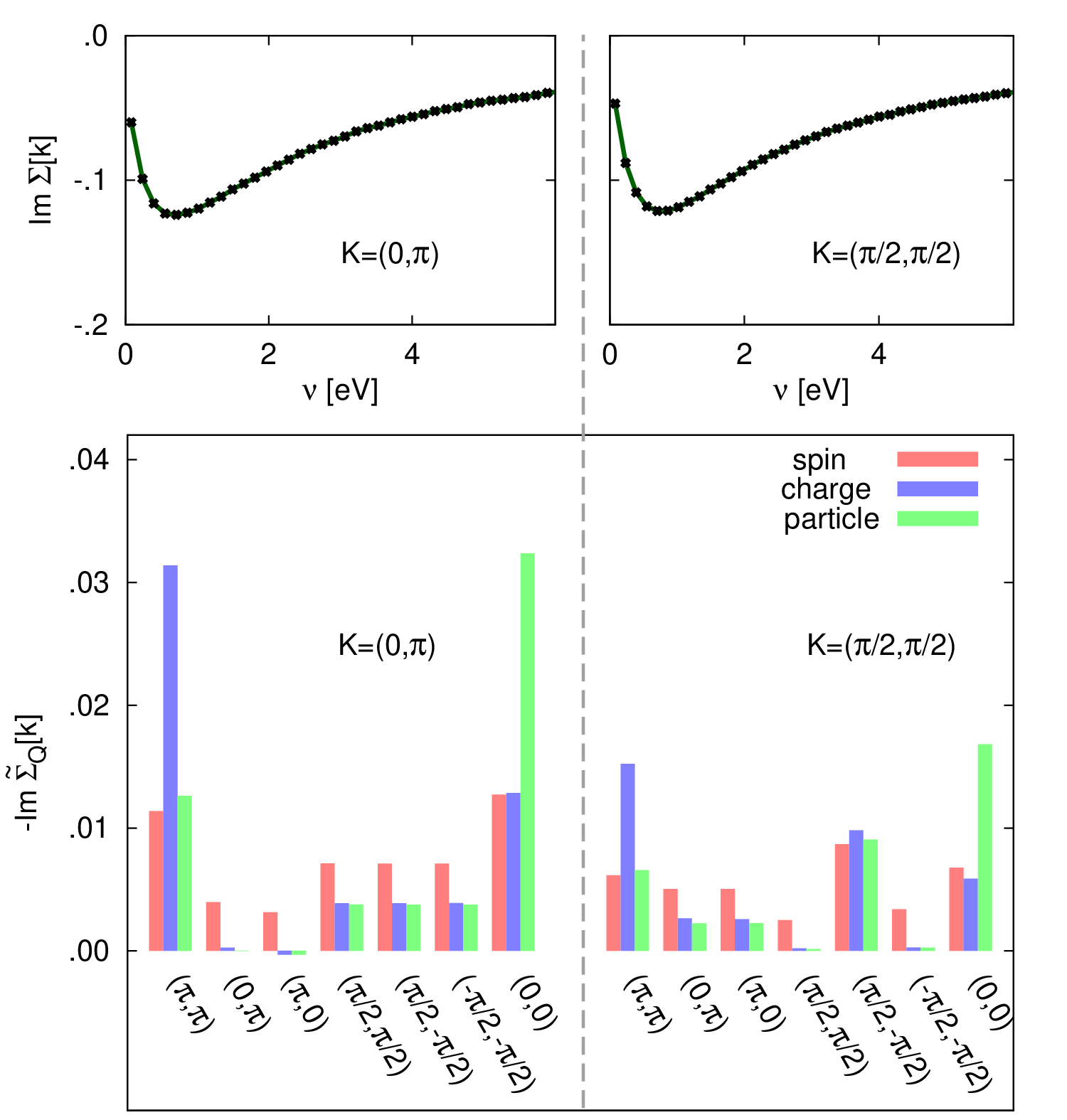}}}} \vskip 3mm
\includegraphics[width=8cm]{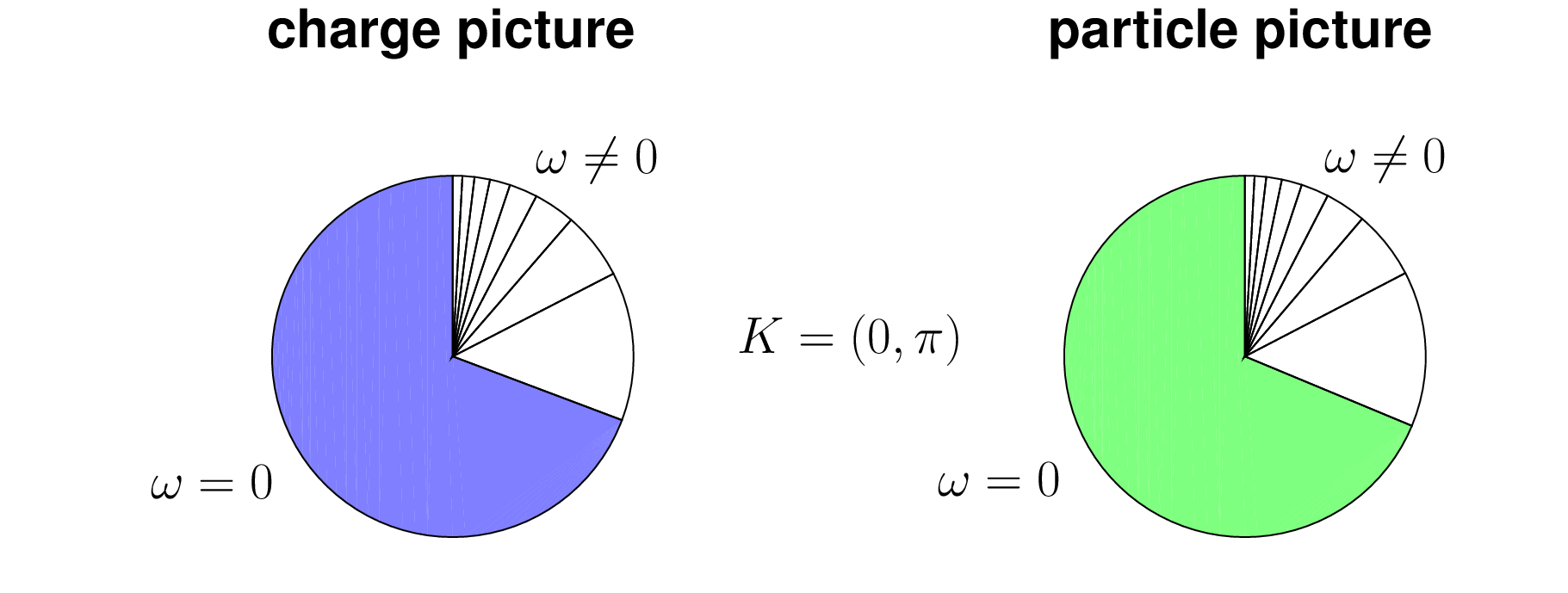}
\caption{\label{fig1} (Color online) Fluctuation diagnostics of Im $\Sigma({\bf K},\nu)$ (first row) for the attractive Hubbard model. The histogram shows the contributions of Im $\widetilde{\Sigma}_{\bf Q}({\bf K},\pi/\beta)$  from different values of ${\bf Q}$  in the spin, charge and particle-particle representations for the attractive 2D Hubbard model (see text). The pie charts display the relative magnitudes of $\lvert \mbox{Im} \widetilde{\Sigma}_{\omega}(\mathbf{K},\pi/\beta)\rvert$ for the first eight Matsubara frequencies $|\omega|$ in the charge and particle-particle picture, respectively.}
\end{figure} Inserting these results in Eq.~(\ref{eq:1}) and performing variable transformations, we recover Eq.~(\ref{eq:1}), with $F_{\uparrow \downarrow}$ replaced by $F_{sp}$, $F_{ch}$  or $F_{pp}$. These three expressions,

{\small
\begin{eqnarray} 
\Sigma(k)-\Sigma_{\text{H}}  &  & =    
 \frac{U}{\beta^2 N}\sum_{k',q} \, F_{sp}(k,k'; q) \, g(k')g(k'\! +\!q)g(k \!+ \!q),  \\
       =  &  & - \, 
 \frac{U}{\beta^2 N}\sum_{k',q} \, F_{ch}(k,k'; q) \, g(k')g(k'\!+\!q)g(k\!+\!q),  \\
            = & &  - \,
  \frac{U}{\beta^2 N}\sum_{k',q} \, F_{pp}(k,k';q) \, g(k')g(q\!-\!k')g(q\!-\!k), 
\end{eqnarray}}
\noindent
yield the same result for $\Sigma$ after all internal summations are performed ($\Sigma_{\text{H}}$ denotes the constant Hartree term $\frac{Un}{2}$). Crucial physical insight can be gained at this stage, by performing {\sl partial} summations. We can, e.g., perform all summations, {\sl except} for the one over the transfer momentum {\bf Q}. This gives $\widetilde{\Sigma}_{\bf Q}(k)$, i.e. the contribution to $\Sigma$ for fixed ${\bf Q}$, so that $\Sigma(k)=\sum_{\bf Q}\widetilde{\Sigma}_{\bf Q}(k)$. The vector ${\bf Q}$ corresponds to a specific spatial pattern given by the Fourier factor $e^{i\mathbf{Q}\mathbf{R}_i}$. For a given representation such a spatial structure is associated to a specific collective mode, e.g., ${\bf Q}\!=\!(\pi,\pi)$ for antiferromagnetic or charge-density-wave (CDW) and ${\bf Q} =(0,0)$ for superconducting or ferromagnetic fluctuations. Hence, if one of these contributions dominates, $\widetilde{\Sigma}_{\bf Q}(k)$ is strongly peaked at the ${\bf Q}$-vector of that collective mode, {\sl provided that} the corresponding representation of the EOM is used. On the other hand, in a different representation, {\sl not} appropriate for the dominant mode $\widetilde{\Sigma}_{\bf Q}(k)$ will display a weak ${\bf Q}$ dependence. These heuristic considerations can be formalized by expressing $F$ through its main momentum and frequency structures \cite{RVT}, see Supplementary section. Hence, in cases where the impact of the different fluctuation channels on $\Sigma$ is not known {\sl a priori}, the analysis of the  ${\bf Q}$-dependence of  $\widetilde{\Sigma}_{\bf Q}(k)$ in the alternative representations of the EOM will provide the desired diagnostics. Below, we show that this procedure works well for the cases of the 2D (attractive and repulsive) Hubbard models, allowing for an interpretation of the origin of the pseudogap phases observed there.

{\sl Results for the attractive Hubbard model.} -- To demonstrate the applicability of the fluctuation diagnostics, we start from a case where the underlying, dominant physics is well understood, namely the attractive Hubbard model,  $U\!<\!0$ . This model captures the basic mechanisms of the BCS/Bose-Einstein crossover \cite{BCSBE,BCSBE2,BCSBE3,BCSBE4,BCSBE5} and has been intensively studied  both analytically and numerically, e.g.,  with QMC \cite{Moreo1991,Kyung2001,Paiva2004} and DMFT \cite{DMFTUneg,DMFTUneg2,DMFTUneg3,DMFTUneg4}. Because of the local attractive interaction, the dominant collective modes are necessarily $s-$wave pairing fluctuations [${\bf Q} =(0,0)$] in the particle channel, and, for filling $n \sim 1$, CDW fluctuations [${\bf Q} = (\pi,\pi)$] in the charge channel. As we show in the following, this underlying physics is well captured by our fluctuation diagnostics.

We present here our  DCA results computed on a cluster with $N_c=8$ sites for a 2D Hubbard model with the following parameter set: $t=-0.25$ eV,  $U=-1$ eV, $\mu=-0.53$ eV and $\beta=40$ eV$^{-1}$. This leads to the occupancy  $n=0.87$, for which, at this $T$, no  superconducting long-range order is observed in DCA.   The lower panels of Fig.~\ref{fig1} show the fluctuation diagnostics for $\Sigma$. The histogram depicts the different contributions to Im $\Sigma[{\bf K},\nu]$ for ${\bf K}=(0,\pi)$ and $(\pi/2,\pi/2)$ (upper panel of Fig.\ \ref{fig1}) at the lowest Matsubara frequency ($\nu= \pi/\beta$) as a function of the momentum transfer ${\bf Q}$ within the three representations [spin, charge and particle, i.e., via Eqs. (2), (3), (4)]. We observe large contributions for ${\bf Q}=(\pi,\pi)$ in the charge representation (blue bars) {\sl and} for ${\bf Q}=(0,0)$ in the particle-particle representation (green bars). At the same time,  {\sl no}  ${\bf Q}$ dominates in the spin picture. Hence, the fluctuation diagnostics correctly identifies the key role of CDW and $s-$wave pairing  fluctuations in this system. This outcome is supported by a complementary analysis in frequency space (pie-chart in Fig.~\ref{fig1}): Defining $\widetilde{\Sigma}_{\omega}(\mathbf{K},\nu)$ as contribution to the self-energy where in Eq.\ (1) all summations, {\sl except} the one over the transfer frequency $\omega$ are performed, we observe a largely dominant contribution at $\omega=0$ ($\sim 70\%$) both in the charge and particle-particle pictures. This proves that the corresponding fluctuations are well-defined and long-lived. 

%This clear-cut outcome supports the applicability of our analyis to other, less defined cases, where the physical interpretation  of important spectral features (e.g., pseudogaps, kinks, waterfalls, etc.) might appear compatible with different or even conflicting theoretical scenarios. 

\begin{figure}
{\rotatebox{-90}{\resizebox{9.0cm}{!}{\includegraphics {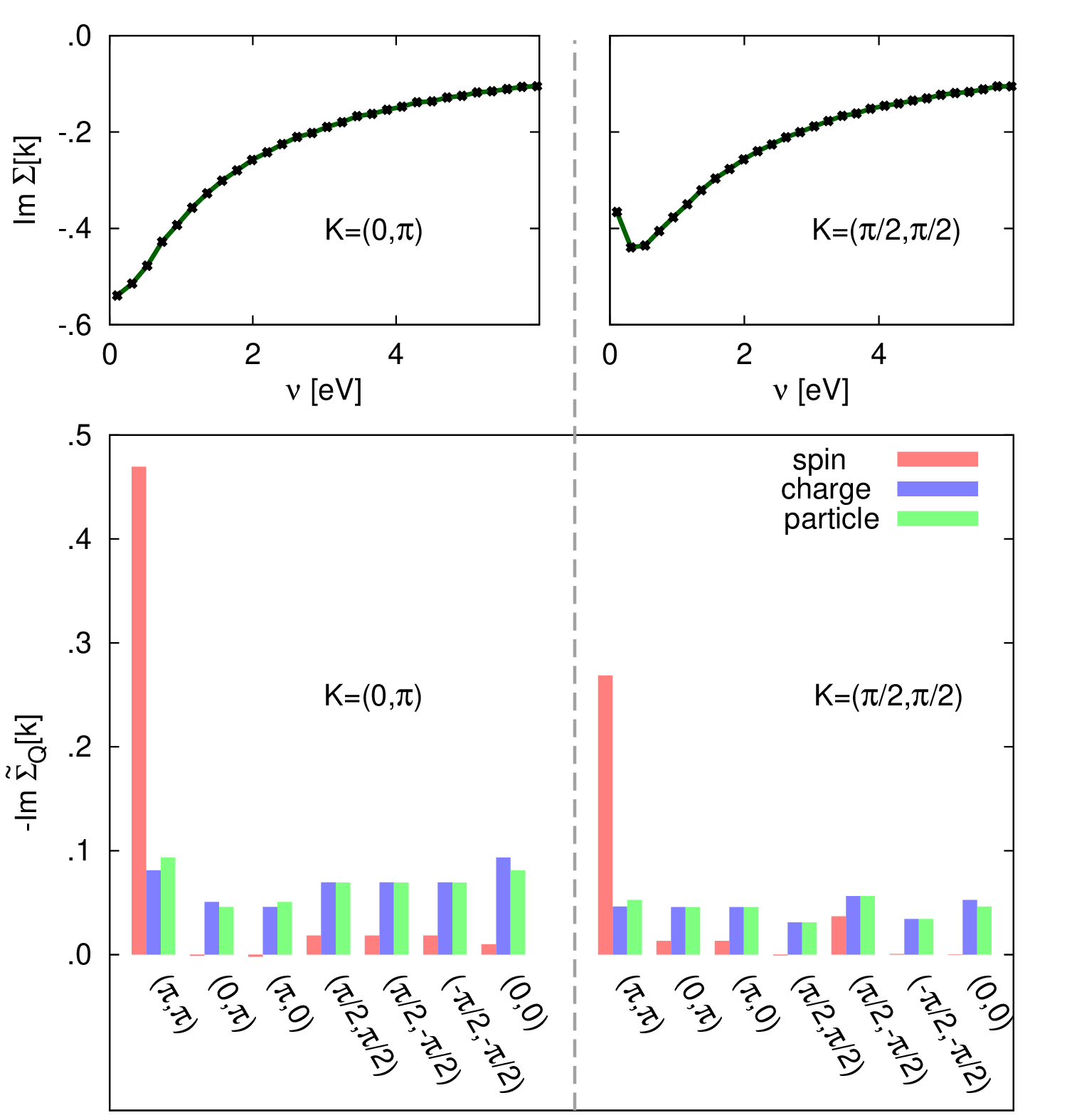}}}}
\vskip 3mm
\includegraphics[width=8cm]{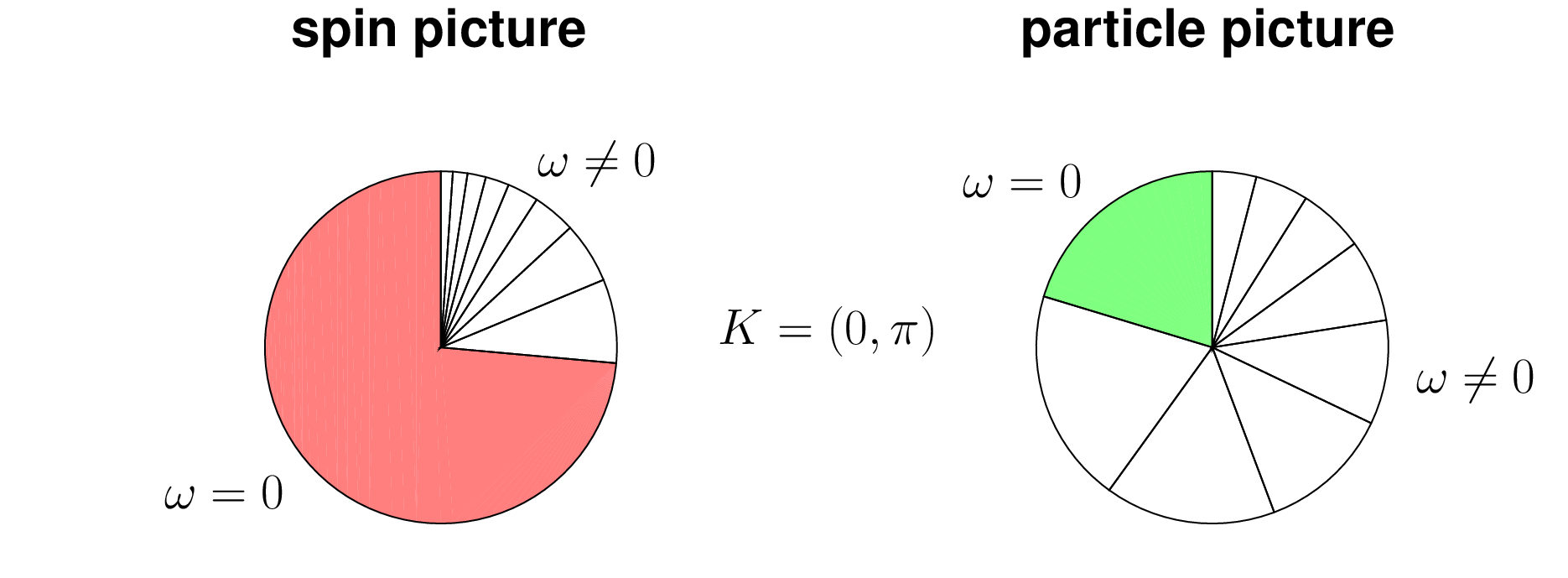}
\caption{\label{fig2} (Color online) As for Fig.\ 1: Fluctuation diagnostics of the electronic self-energy, for the case of the repulsive Hubbard model (see text).}
\end{figure}

{\sl Results for the repulsive Hubbard model.} -- We now apply the fluctuation diagnostics to the much more debated physics of the repulsive Hubbard model in 2D, focusing on the analysis of the pseudogap regime. As before, we use  DCA calculations with a cluster of $N_c =8$ sites. $\Sigma$ and $F$ have been calculated using the Hirsch-Fye \cite{HirschFye} and Continuous Time \cite{CTAUX,CT} QMC methods, accurately cross-checking the results. Specifically, we consider the parameter set $t=-0.25$ eV, $U=1.6$ eV, $\mu=0.8$ eV (corresponding to $n=1$) and $\beta=30$ eV$^{-1}$. For these parameters, the electronic self-energy (see upper panels of Fig. \ref{fig2}) displays strong momentum differentiation between the ``antinodal'' [${\bf K}=(0,\pi)$] and the ``nodal'' [${\bf K}=(\pi/2,\pi/2)$] momentum, with a clear pseudogap behavior at the antinode \cite{Gull2010,pseudogap}. 

The fluctuation diagnostics is performed in Fig.~\ref{fig2}, where we show the contributions to Im $\Sigma[{\bf K},\pi/\beta]$ for ${\bf K}=(0,\pi)$ and $(\pi/2,\pi/2)$ (upper panels) as a function of the transfer momentum ${\bf Q}$ in the three representations. This illustrates clearly the underlying physics of the pseudogap. In the {\sl spin} representation (red bars in the histogram), the ${\bf Q}=(\pi,\pi)$ contribution dominates, and contributes more than $80 \%$ and $70 \%$ of the result for ${\bf K}=(0,\pi)$ and ${\bf K}=(\pi/2,\pi/2)$, respectively. Conversely, {\sl all} the contributions at other transfer momenta ${\bf Q} \neq (\pi,\pi)$ are about an order of magnitude smaller. The dominant ${\bf Q}=(\pi,\pi)$-contribution is also responsible for the large momentum differentiation, being  almost {\sl twice} as large for the antinodal self-energy. Performing the same analysis in the {\sl charge} (blue bars) or {\sl particle-particle} (green bars) representation, we get a completely different shape of the histogram. In both cases, the contributions to $\Sigma$ are almost uniformly distributed among all transfer momenta  ${\bf Q}$.

On the basis of our previous considerations, we do not find important contributions to $\Sigma$ from {\sl charge} or {\sl pairing} modes, while the histogram in the spin-representation marks the strong impact of antiferromagnetic fluctuations \cite{spinflu1, spinflu2,spinflu3,spinflu4, spinflu5, Gull2009, Gull2012}. This picture is further supported by the complementary frequency analysis. The pie chart in Fig.\ \ref{fig2} is dominated by the $\omega=0$ contribution in the spin picture, reflecting the long-lived nature of well-defined spin-fluctuations. At the same time, in the particle (and charge, not shown) representation, the contributions are more uniformly distributed among all $\omega$'s, which corresponds to short-lived pairing (charge) fluctuations. 

{\sl Physical interpretation of the pseudogap.}-- We are now in the position to draw some general conclusions on the physics underlying a pseudogap. These considerations are relevant for the underdoped cuprates, up to the extent their low-energy physics is captured by the $8$-site DCA for the repulsive 2D Hubbard model. For simplicity, we focus here on our data for the unfrustrated model at half-filling, which exhibits a pseudogap in the parameter regime considered. By means of fluctuations diagnostics we identify a well-defined [$\mathbf{Q}\!=\!(\pi,\pi)$] collective
spin-mode to be responsible (on the 75\% level) both for the momentum differentiation of $\Sigma$ and for its pseudogap behavior at the antinode: The large values of $\widetilde{\Sigma}_{\mathbf{Q}}$ at $\mathbf{Q}\!=\!(\pi,\pi)$ and $\widetilde{\Sigma}_{\omega}$ at $\omega=0$ are the distinctive hallmark of {\sl long-lived} and {\sl extended} (antiferromagnetic) spin-fluctuations. At the same time, the rather uniform $\mathbf{Q}$- and $\omega$-distribution of $\widetilde{\Sigma}_{\mathbf{Q}}$ and $\widetilde{\Sigma}_{\omega}$ in the charge/particle pictures shows that the well-defined spin mode can be {\sl also} viewed as {\sl short-lived} and {\sl short-range} charge/pair fluctuations.  The latter cannot be interpreted, hence, in terms of {\sl preformed} pairs. This scenario for the pseudogap matches very well the different estimates of fluctuation strengths in previous DCA studies \cite{pseudogap,pseudogap1,Gull2012}. We also emphasize the general applicability of our result (see Supplementary): A well defined mode in one channel appears as short-lived fluctuations in other channels. This dichotomy is not visible anymore in $\Sigma$, which makes our fluctuations diagnostics a powerful tool for identifying the {\sl most convenient} viewpoint to understand the physics responsible of the observed spectral properties. 

Let us finally turn our attention to the still open question about the
impact of superconducting $d$-wave fluctuations on the normal-state spectra in the pseudogap regime 
of the Hubbard model. The instantaneous fluctuations are defined as $\langle
\Delta_d^{\dagger}\Delta_d\rangle$, with
$\Delta_d^{\dagger}=\sum_{\bf K}f({\bf K})c^{\dagger}_{{\bf
K}\uparrow}c^{\dagger}_{-{\bf K}\downarrow}$
and $f({\bf K})={\rm cos} K_x-{\rm cos}K_y$. These ${\bf Q}=0$ fluctuations are certainly strong in proximity of the superconducting phase, but they were also found\cite{pseudogap} to be significant over short distances in the pseudogap regime. Their intensity gets stronger as $U$ is increased, beyond the values where superconductivity exists.
The  expression for $\tilde \Sigma_{{\bf Q}=(0,0)}$ in the
particle picture
is closely related to $\langle \Delta_d^{\dagger}\Delta_d\rangle$,
except that the
factor $f({\bf K})$ is missing in $\tilde \Sigma_{\bf Q}$ (see
Supplementary). One might therefore have expected that large ${\bf Q=0}$  pair fluctuations, irrespectively of their lifetime, would have contributed strongly to $\Sigma$. For unconventional superconductivity, e.g., $d-$wave, this does not happen. The reason is the angular variation of $f({\bf K})$. For strong pair fluctuations, the variations of $f({\bf K})$ make the contributions to the fluctuations add up, while the contributions to $\Sigma$ then tend to cancel. This explains why suppressing superconductivity fluctuations \cite{pseudogap,pseudogap1,Vilk1997,Sigmak,DF1, DGA1, DGA2, Avella2014} does not affect the description of the pseudogap of the Hubbard model. In the case of a purely local interaction such as in the EOM like Eq.~(\ref{eq:1}), enhanced $\langle \Delta_d^\dagger \Delta_d \rangle$ fluctuations are mostly averaged out by the momentum summation (see Supplementary). 

%Obviously, the latter conclusion might change if the {\sl non-local} electrostatic interactions become important.

{\sl Conclusions.} -- We have shown that if a simultaneous calculation of the self-energy and the vertex functions is performed, it is possible to identify the impact of the different collective modes on the spectra of correlated systems (``fluctuation diagnostics''). This  is achieved by expressing the equation of motion for $\Sigma$ in different representations (e.g., spin/charge/particle), which avoids all the intrinsic instabilities of parquet decompositions. We apply this procedure to the $U<0$ and $U>0$ 2D Hubbard model. In the attractive case we have confirmed the dominant role of pair fluctuations, supporting the validity of our approach. For the repulsive model, relevant for the physics of the underdoped cuprates, spin fluctuations emerged as mainly responsible for the spectral function results, in agreement with other studies \cite{spinflu1, spinflu2,spinflu3, spinflu4, spinflu5, Gull2012}.  The same well-defined spin modes might appear, on a different perspective, as strong, but rapidly decaying,  pair fluctuations. Finally, for a purely local interaction, $d-$wave pairing fluctuations will only weakly affect the pseudogap spectral properties even on the verge of the superconducting transition. 

These results, as well as the insight on the pseudogap physics, suggest that fluctuation diagnostics can be broadly used in future studies. The progress in calculating vertex functions \cite{Gunnarsson2010, RVT, Hafermann2014} will allow its applicability also to other, more complex, multi-orbital models \cite{LDACDMFT1,LDAVCA,LDACDMFT2,LDADCA,abinitioDGA,LDAFRG,multiDCA}: Here, due to the increased number of degrees of freedom, the identification of the dominant fluctuation mode(s) will be of the utmost importance for a correct physical understanding. 

{\sl Acknowledgments.} --  We thank A.\ Tagliavini, C.\ Taranto, S. Andergassen, and M.\ Capone for insightful discussions. We acknowledge support from the FWF through the PhD School ``Building Solids for Function''  (TS, Proj.\ Nr.\ W1243) and the project I-610 (GR, AT), from the research unit FOR 1346 of the DFG (GS), from  MINECO: MAT2012-37263-C02-01 (JM), and from the Simons foundation (JPFL,EG). GS and AT also acknowledge the hospitality in Campello sul Clitunno.

\end{document}

\title{Fluctuation diagnostics of the electron self-energy: Origin of the pseudogap physics\\
--Supplementary Material--}                                         
%\title{Physical interpretation of electron self-energy in Quantum Monte-Carlo calculations}
%\title{Physical interpretation of electron self-energy via equation of motion and parquet decomposition calculations}
\author{O.~Gunnarsson, T.~Sch\"afer, J.\ P.\ F.~LeBlanc, E.~Gull, 
 J.~Merino, G.~Sangiovanni, G.~Rohringer, and A.~Toschi}

\date{\today}

\maketitle

\section{Definitions and important relations}

In this section we define the model and the central physical quantities we use throughout the manuscript. The Hamiltonian of the two-dimensional Hubbard model considered here reads
\begin{equation}
 \label{supp:hamilt}
 \hat{\mathcal{H}}=t\sum_{\langle ij \rangle,\sigma}\hat{c}^{\dagger}_{i\sigma}\hat{c}_{j\sigma}+U\sum_i\hat{n}_{i\uparrow}\hat{n}_{i\downarrow},
\end{equation}
where $\hat{c}^{(\dagger)}_{i\sigma}$ (creates) annihilates an electron with spin $\sigma$ at the lattice site $i$ (corresponding to a lattice vector $\mathbf{R}_i$), $\hat{n}_{i\sigma}=\hat{c}^{\dagger}_{i\sigma}\hat{c}_{i\sigma}$. $t$ describes the hopping amplitude between neighboring sites and $U$ is the Coulomb interaction between electrons at the same lattice site. In the manuscript we consider the paramagnetic phase with $n$ particles per lattice site at a given temperature $T=1/\beta$.

The central quantities we are dealing with throughout the manuscript are the one- and two-particle Green's functions $G_1$ and $G_2$ as well as their irreducible parts. The $n$-particle Green's function is generally defined as the time-ordered product of $n$ creation and $n$ annihilation operators:
\begin{align}
\label{equ:defgreen}
 &G_{n,(i_1\sigma_1)\ldots (i_{2n}\sigma_{2n})}(\tau_1,\ldots,\tau_{2n})=\\ \nonumber&\hspace{1cm}=\left\langle T\left(\hat{c}^{\dagger}_{i_1\sigma_1}(\tau_1)\hat{c}_{i_2\sigma_2}(\tau_2)\ldots\hat{c}_{i_{2n}\sigma_{2n}}(\tau_{2n})\right)\right\rangle,
\end{align}
where $T$ is the time-ordering operator and $\hat{c}^{(\dagger)}_{i_j\sigma_j}(\tau_j)$ represents a (creation) annihilation operator in the Heisenberg picture at time $\tau_j$. $\langle\ldots\rangle\;\widehat{=}\;\frac{1}{Z}\mbox{Tr}[e^{-\beta(\hat{\mathcal{H}}-\mu\hat{\mathcal{N}})}\ldots]$ with the total particle number operator $\hat{\mathcal{N}}\!=\!\sum_{i\sigma}\hat{n}_{i\sigma}$ and the chemical potential $\mu$. $Z=\mbox{Tr}[e^{-\beta(\hat{\mathcal{H}}-\mu\hat{\mathcal{N}})}]$ is the grandcanonical partition function. Performing a Fourier-transform with respect to the time ($\tau_j\rightarrow\nu_j$) and space ($\mathbf{R}_{i_j}\rightarrow\mathbf{K}_j$) variables, time and lattice translational invariance (i.e., energy and crystal momentum conservation) allow for a restriction to $2n-1$ (fermionic) Matsubara frequencies and $\mathbf{k}$-vectors \cite{RVT}. Moreover, in the paramagnetic phase considered here, SU(2) symmetry leads to a tremendous simplification of the spin-dependence of the Green's functions \cite{RVT,georgthesis}. Hence, the Fourier representations of the relevant one- and two-particle Green's function are given by
\begin{widetext}
\begin{align}
 \label{equ:deffourgreen1}
 g(k)&=\int_0^{\beta}d\tau\sum_ie^{-i[\nu\tau-\mathbf{K}\cdot\mathbf{R}_i]}G_{1,(i\sigma )(i_0\sigma)}(\tau,0) \\ \label{equ:effourgreen2}
 g_{2,\sigma\sigma'}(k,k',q)&=\int_0^{\beta}d\tau_1d\tau_2d\tau_3\sum_{i_1i_2i_3}e^{-i[\nu\tau_1-\mathbf{K}\cdot\mathbf{R}_{i_1}]}e^{i[(\nu+\omega)\tau_2-(\mathbf{K}+\mathbf{Q})\cdot\mathbf{R}_{i_2}]}e^{-i[(\nu'+\omega)\tau_3-(\mathbf{K'}+\mathbf{Q})\cdot\mathbf{R}_{i_3}]} \times\\[-0.5cm]\nonumber &\hspace{8.5cm} G_{2,(i_1\sigma)(i_2\sigma)(i_3\sigma')(i_0\sigma')}(\tau_1,\tau_2,\tau_3,0),
\end{align}
\end{widetext}
where the index $i_0$ denotes the lattice site at the origin of the coordinate system, i.e., $\mathbf{R}_{i_0}=(0,0)$. On the left hand side of these equations we have adopted the four-vector notation $k^{(\prime)}\widehat{=}(\nu^{(\prime)},\mathbf{K^{(\prime)}})$ for the fermionic Matsubara frequencies $\nu^{(\prime)}$ and $q\widehat{=}(\omega,\mathbf{Q})$ for the (transferred) bosonic Matsubara frequency $\omega$. As usual, we represent two-particle quantities as a function of two fermionic ($\nu,\nu'$) and one bosonic ($\omega$) Matsubara frequencies. Let us also point out that $g_{2,\sigma\sigma}$ (i.e., $g_2$ with both spin-arguments taken to be equal) is not an independent quantity since it can be derived from $g_{2,\sigma(-\sigma)}$ by means of SU(2) and crossing symmetry relations \cite{RVT}.

Of special interest for our analysis are the one-particle irreducible (1PI) parts of the one- and the two-particle Green's functions, i.e., the self-energy $\Sigma(k)$ and vertex $F^{kk'q}_{\sigma\sigma'}$. The self-energy is defined via the Dyson equation:
\begin{equation}
 \label{equ:dyson}
 \Sigma(k)=g_0^{-1}(k)-g^{-1}(k)=(i\nu+\mu-\varepsilon_\mathbf{K})-g^{-1}(k),
\end{equation}
where $g_0(k)\!=\!(i\nu+\mu-\varepsilon_{\mathbf{K}})^{-1}$ is the non-interacting Green's function and $\varepsilon_{\mathbf{K}}$ is the dispersion of the non-interacting system, i.e.,  $\varepsilon_{\mathbf{K}}=-2t(\cos K_x+\cos K_y)$ for the 2d Hubbard model.
As for the 1PI vertex function, its definition in terms of the one- and two-particle Green's functions reads:
\begin{widetext}
\begin{align}
 \label{equ:vertex}
 %F_{\sigma\sigma'}&(k,k',q)=-\frac{1}{g(k)g(k+q)g(k')g(k'+q)}\times\\ \nonumber&\hspace{-0.5cm}\left[g_{2,\sigma\sigma'}(k,k',q)-\beta g(k)g(k')\delta_{q0}+\beta g(k)g(k+q)\delta_{kk'}\right],
 F_{\sigma\sigma'}&(k,k',q)=-\frac{1}{g(k)g(k+q)g(k')g(k'+q)}\left[g_{2,\sigma\sigma'}(k,k',q)-\beta g(k)g(k')\delta_{q0}+\beta g(k)g(k+q)\delta_{kk'}\delta_{\sigma\sigma'}\right],
\end{align}
\end{widetext}
It is important to recall that $\Sigma$ and $F$ are related by an exact equation, the so-called Schwinger-Dyson equation of motion (EOM), which reads:
\begin{equation}
 \label{equ:EOM}
 \Sigma(k)=\frac{Un}{2}-\frac{U}{\beta^2N}\sum_{k',q}F_{\uparrow\downarrow}(k,k',q)g(k')g(k'+q)g(k+q),
\end{equation}
where $N$ is the normalization of the momentum summation. Let us also recall that in Eq.\ (\ref{equ:EOM}) $F_{\uparrow\downarrow}$ can be replaced by $\pm F_{ch,sp}$ [with $F_{ch,sp}\!=\! (F_{\uparrow\uparrow}\!\pm\!F_{\uparrow\downarrow})$]  since $\sum_{k',q}F_{\uparrow\uparrow}(k,k',q)g(k')g(k'+q)g(k+q)\!=\!0$ due to the SU(2) symmetry relation $F_{\uparrow\uparrow}(k,k',q)\!=\!F_{\uparrow\downarrow}(k,k',q)\!-\!F_{\uparrow\downarrow}(k,k+q,k'-k)$, see Refs. \cite{RVT,georgthesis}. Moreover $F_{\uparrow\downarrow}$ in Eq.\ (\ref{equ:EOM}) can be replaced by its particle-particle representation $F_{pp}(k,k',q)\!=\!F_{\uparrow\downarrow}(k,k',q-k-k')$.

\section{Fluctuation decomposition of the vertex}

The physical interpretation of our numerical results, as presented in our manuscript, is supported by a precise analytical derivation valid for the weak-coupling regime. Specifically we will consider in the following an approximation for the vertex function $F_r$, $r=ch,sp,pp$, entering in the EOM [Eq.(1) in the manuscript]. In this approximation, we retain all principal frequency and momentum structures of the vertex functions, i.e. (beyond the bare interaction $U$) the main and secondary diagonal and the constant background, see Ref.\ \cite{RVT,Karrasch}. Physically, these main features of $F$ correspond to the different susceptibilities (response-functions) $\chi_r(\mathbf{Q},\omega)$, $r=ch,sp,pp$. They can be calculated as
\begin{subequations}
\label{equ:susc} 
\begin{align}
 \label{equ:suscc}
 \chi_{ch}(q)&=\frac{1}{\beta^2N^2}\sum_{k,k'}\left[g_{2,\uparrow\uparrow}(k,k',q)+g_{2,\uparrow\downarrow}(k,k',q)-\right.\\[-0.3cm] &\hspace{4.5cm}\left.\nonumber 2\beta  g(k)g(k')\delta_{q0}\right]\\
 \label{equ:suscm}
 \chi_{sp}(q)&=\frac{1}{\beta^2N^2}\sum_{k,k'}\left[g_{2,\uparrow\uparrow}(k,k',q)-g_{2,\uparrow\downarrow}(k,k',q)\right]\\
 \label{equ:suscpp}
 \chi_{pp}(q)&=\frac{1}{\beta^2N^2}\sum_{k,k'}g_{2,\uparrow\downarrow}(k,k',q-k'-k),
\end{align}
\end{subequations}
where $N$ is the normalization of the momentum summation and $\chi_{pp}(q)$ defines the {\sl $s$-wave} particle-particle susceptibility. In our approximation the vertex $F$ will be now expressed as
\begin{widetext}
\begin{subequations}
\label{equ:fasympt}
\begin{align}
&F_{ch}(k,k',q)\approx U+U^2\left[-\chi_{ch}(q)+\frac{3}{2}\chi_{sp}(k'-k)+\frac{1}{2}\chi_{ch}(k'-k)-\chi_{pp}(k+k'+q)\right]\label{equ:fasymptcharge}\\
&F_{sp}(k,k',q)\approx -U -U^2\left[\chi_{sp}(q)+\frac{1}{2}\chi_{sp}(k'-k)-\frac{1}{2}\chi_{ch}(k'-k)+\chi_{pp}(k+k'+q)\right]\label{equ:fasymptspin}\\
&F_{pp}(k,k',q)\approx U+U^2\left[-\frac{1}{2}\chi_{ch}(q-k-k')+\frac{1}{2}\chi_{sp}(q-k-k')+\chi_{sp}(k'-k)-\chi_{pp}(q)\right]\label{equ:fasymptpp}
\end{align}
\end{subequations}
\end{widetext}
As mentioned above, such an approximation can be rigorously justified only in the weak-coupling regime, i.e., for small interaction values $U$ ($U\!\ll\!t$). In this limit, however, Eqs.~(\ref{equ:fasympt}) allow for a immediate understanding of how different fluctuations contribute to the self-energy. In this respect, let us recall that each susceptibility $\chi_r(\mathbf{Q},\omega)$ has a clear physical meaning: They describe the (linear) response of the system with respect to an external forcing field, which is associated to the specific channel $r$ ($r\!=\!ch$ $\rightarrow$ chemical potential, $r\!=\!sp$ $\rightarrow$ (staggered) magnetic field, $r\!=\!pp$ $\rightarrow$ pairing field). They become obviously very large in the vicinity of a corresponding second order phase transition. More specifically, the {\sl static} susceptibility $\chi_r(\mathbf{Q},\omega=0)$ gets strongly enhanced at a specific momentum, $\mathbf{Q_0}$, if the system exhibits large fluctuations in the channel $r$, which are associated with the spatial pattern defined by $e^{i\mathbf{Q_0}\cdot\mathbf{R}_i}$ (see discussion in the main text). Hence, we generally expect that the susceptibilities $\chi_r(q)$ (at $\omega\!=\!0$ and $\mathbf{Q}\!=\!\mathbf{Q_0}$) yield the most relevant contributions to the self-energy, if the system exhibits large fluctuations in the corresponding channel(s) $r$ [see Eqs.~(\ref{equ:fasympt}) and the EOM]. In the following, we will apply these considerations to the cases of the repulsive ($U\!>\!0$) and the attractive ($U\!<\!0$) Hubbard model discussed in the main text.

Let us assume that in the {\sl repulsive Hubbard model} antiferromagnetic [$\mathbf{Q}\!=\!\mathbf{\Pi},\mathbf{\Pi}\!=\!(\pi,\pi)$] spin-fluctuations dominate (this is most likely the case at half-filling). We will analyze in the following how --in this situation-- the different frequency ($\widetilde{\Sigma}_{\omega}$) and momentum ($\widetilde{\Sigma}_{\mathbf{Q}}$) contributions to the self-energy, as depicted in our histograms/pie charts for the self-energy decomposition in Fig.\ 2, are interpreted in terms of the approximate form for the vertex $F_r$ in Eqs.~(\ref{equ:fasympt}). In a spin dominated situation, the most relevant contributions to $F_r$ will originate from $\chi_{sp}(\mathbf{Q}\!=\!\mathbf{\Pi},\omega\!=\!0)$. Following the above considerations and replacing the exact vertex functions $F_r$ by their approximate forms (\ref{equ:fasympt}) in the calculation of $\widetilde{\Sigma}_{\mathbf{Q}}$ and $\widetilde{\Sigma}_{\omega}$ we, hence, arrive at the following conclusions: 

In the {\bf{\color[rgb]{0.7,0,0}  spin}-picture}, $\chi_{sp}$ appears as a function of $\mathbf{Q}$ and $\omega$ in $F_{sp}$ [see Eqs.~(\ref{equ:fasymptspin})], independent of $k'$. In this situation {\sl each} term in the $k'$-sum in the EOM includes the large contribution $\chi_{sp}(\mathbf{Q}\!=\!\mathbf{\Pi},\omega\!=\!0)$ to $\widetilde{\Sigma}_{\mathbf{Q}}$ and $\widetilde{\Sigma}_{\omega}$. On the other hand, for $\mathbf{Q}\!\ne\!\mathbf{\Pi}$ or $\omega\!\ne\!0$ the largest contribution to the $k'$ summation stems from the {\sl single} term proportional to $\chi_{sp}(k'-k)$ in Eq. (\ref{equ:fasymptspin}), evaluated for $(\mathbf{K'}\!-\!\mathbf{K})\!=\!\mathbf{\Pi}$ and $\nu'\!-\!\nu\!=\!0$. This explains the rather small values of $\widetilde{\Sigma}_{\mathbf{Q}}$ and $\widetilde{\Sigma}_{\omega}$ for $\mathbf{Q}\ne\mathbf{\Pi}$ or $\omega\ne0$, respectively, in the spin picture. We note that this situation corresponds to histograms and pie charts, very similar to those observed for the DCA calculation of Fig. 1 in the manuscript. 

At the same time, in the {\bf{\color[rgb]{0,0,0.7} charge}} {\sl and} {\bf{\color[rgb]{0,0.5,0} particle-particle} representations}, $\chi_{sp}$ appears only as a function of $k'\!-\!k$ (or $q\!-\!k\!-\!k'$), see Eqs.~(\ref{equ:fasymptcharge}) and (\ref{equ:fasymptpp}). Therefore, when performing the partial summations over $k'$ in the EOM, only the {\sl single} contribution for $\mathbf{K}'\!-\!\mathbf{K}\!=\!\mathbf{\Pi}$ and $\nu'\!-\!\nu\!=\!0$ ($\mathbf{Q}\!-\!\mathbf{Q}\!-\!\mathbf{K}'\!=\!\mathbf{\Pi}$ and $\nu\!-\!\nu'\!-\!\omega\!=\!0$) is large in this sum. On the other hand, such a contribution appears for {\sl each} value of $\mathbf{Q}$ and $\omega$. This explains well the fact that in the charge and the particle-particle pictures the contributions $\widetilde{\Sigma}_{\mathbf{Q}}$ and $\widetilde{\Sigma}_{\omega}$, respectively, to the self-energy are uniformly distributed among all values of $\mathbf{Q}$ and $\omega$ as it is observed in the histograms and pie chart (only particle-particle) in Fig.\ 2. Let us stress, that $\chi_{sp}$ {\sl does} contribute to $F_r$ in the charge and particle-particle picture, but only as a function of $k'\!-\!k$ rather than $q$. Hence, one can argue that in the charge and particle-particle representation spin fluctuations are seen from a not ``convenient'' perspective. From this specific point of view, a well-defined collective spin-mode will appear as short-range (or even local) and short-lived charge or particle-particle fluctuations, as indicated by the democratic distribution of $\widetilde{\Sigma}_{\mathbf{Q}}$ and $\widetilde{\Sigma}_{\omega}$ among all values of $\mathbf{Q}$ and $\omega$. 

Obviously the above analysis is applicable also to the attractive Hubbard model ($U<0$): In this situation charge and particle-particle fluctuations are expected to dominate while spin fluctuations are strongly suppressed. Hence, $\chi_{ch}(\mathbf{Q}\!=\!\mathbf{\Pi},\omega\!=\!0)$ and $\chi_{pp}(\mathbf{Q}\!=\!\boldsymbol{0},\omega=0)$, $\mathbf{0}\!=\!(0,0)$, are enhanced. In the {\bf{\color{red} spin} picture} these arguments for $\chi_{ch}$ and $\chi_{pp}$ appear for {\sl only one} value of $k'$ when performing the $k'$ summation. On the contrary, in the {\bf{\color[rgb]{0,0,0.7} charge}} and in the {\bf{\color[rgb]{0,0.5,0} particle-particle} picture}, $\chi_{ch}$ (or $\chi_{pp}$) is a function of $\mathbf{Q}$ and $\omega$ and the above mentioned large contribution to $\widetilde{\Sigma}_{\mathbf{Q}}$ and $\widetilde{\Sigma}_{\omega}$ appears for {\sl each} value of $k'$. Hence, $\widetilde{\Sigma}_{\mathbf{Q}}$ and $\widetilde{\Sigma}_{\omega}$ get strongly peaked at $\mathbf{Q}\!=\!\mathbf{\Pi}$ and $\omega\!=\!0$, respectively, in the charge description and $\mathbf{Q}\!=\!\boldsymbol{0}$ and $\omega\!=\!0$ in the particle-particle description, while in the spin picture  $\widetilde{\Sigma}_\mathbf{Q}$ is almost independent of $\mathbf{Q}$. 

The above discussion based on the vertex decomposition in Eqs.~(\ref{equ:fasympt}) is rigorously justified only for small values of $U$ where corrections beyond Eqs.~(\ref{equ:fasympt}) are negligible. This highlights the importance of the fluctuation diagnostics approach which is applicable for all values of the interaction. In fact, the fluctuation diagnostics for the DCA self-energy in the pseudogap regime of the repulsive two-dimensional Hubbard model gives gives histograms/pie charts for $\widetilde{\Sigma}_{\mathbf{Q}}$ and $\widetilde{\Sigma}_{\omega}$ dominated by $\mathbf{Q}\!=\!\mathbf{\Pi}$ and $\omega\!=\!0$ in the spin representation, indicating the dominant role played by a well defined and long-lived ($\mathbf{Q}\!=\!\mathbf{\Pi}$, $\omega\!=\!0$) spin collective mode. We note that this hold even in a regime, where Eqs.~(\ref{equ:fasympt}) break down. Specifically we note that while the main bosonic structures of $F$ described in Eqs.~(\ref{equ:fasympt}) give a significant contribution to the self-energy even in the non-perturbative regime, the momentum differentiation observed in the histograms originates from contributions {\sl beyond} Eqs.~(\ref{equ:fasympt}).

\section{${\boldsymbol{d}}$-wave pairing fluctuations}

Let us finally comment on the role of particle-particle fluctuations for the self-energy in the repulsive Hubbard model: In the decomposition of the vertex [Eqs. (\ref{equ:fasympt})] only the $s-$wave superconducting susceptibility $\chi_{pp}$ enters, which is always suppressed for $U\!>\!0$. The corresponding response function for $d-$wave superconductivity, however, is {\sl absent} in the leading structures of the vertex, and, hence, within Eqs.\ (\ref{equ:fasympt}) we do not expect any significant impact of the $d-$wave fluctuations on the self-energy. Obviously in the vicinity to a $d-$wave superconducting instability, the $d-$wave superconducting fluctuation might become relevant for approximating the vertex function \cite{husemann}. However, even in this case, the corresponding ($d-$wave) susceptibility would be strongly enhanced and the fluctuations will cancel in the calculation of the self-energy, as it will be discussed in the next paragraph. 

In the following we show explicitly that $d-$wave pairing fluctuation are irrelevant for the self-energy even close to the corresponding superconducting instability. Considering the $d-$wave pairing operator  $\Delta^{\dagger}\!=\!\sum_{\bf K}f({\bf K})c_{{\bf K}\uparrow}^{\dagger}c_{-{\bf K}\downarrow}^{\dagger}$, where $f({\bf K})\!=f(-{\bf K})= \!{\rm cos} K_x\!-\!{\rm cos}K_y$, the corresponding  pairing fluctuations are given by
\begin{eqnarray}\label{eq:6}
&&\langle \Delta^{\dagger}\Delta\rangle=\sum_{{\bf K},{\bf K}'}f({\bf K})f({\bf K}')\langle c_{{\bf K}\uparrow}^{\dagger}
c_{-{\bf K}\downarrow}^{\dagger}c_{-{\bf K}'\downarrow}^{\phantom \dagger}c_{{\bf K}'\uparrow}^{\phantom \dagger} \rangle \nonumber \\
&& -\sum_{{\bf K}}[f({\bf K})]^2\langle c_{{\bf K}\uparrow}^{\dagger}c_{{\bf K}\uparrow}^{\phantom \dagger}\rangle
\langle c_{-{\bf K}\downarrow}^{\dagger}c_{-{\bf K}\downarrow}^{\phantom \dagger}\rangle,
\end{eqnarray}
To make a connection with the self-energy we rewrite Eq.~(1) of the main text
\begin{eqnarray}\label{eq:5}
&&\frac{N}{U\beta}\sum_{\nu}[\Sigma(k)-\frac{Un}{2}]g(k)=\sum_{{\bf K}',{\bf Q}}
\langle c_{{\bf K}\uparrow}^{\dagger}c_{{\bf Q}-{\bf K}\downarrow}^{\dagger}
c_{-{\bf K}'\downarrow}^{\phantom \dagger}c_{{\bf Q}+{\bf K}'\uparrow}^{\phantom \dagger} \rangle \nonumber \\
&&-\sum_{{\bf K}'}\langle c_{{\bf K}\uparrow}^{\dagger}c_{{\bf K}\uparrow}^{\phantom \dagger}\rangle
\langle c_{{\bf K}'\downarrow}^{\dagger}c_{{\bf K}'\downarrow}^{\phantom \dagger}\rangle,
\end{eqnarray}
where the ${\bf Q}=(0,0)$ term corresponds to contributions from superconductivity fluctuations. For large superconducting fluctuations, one might then have expected also a large contribution to 
$\Sigma$ due to the similar structure of Eq.~(\ref{eq:5}) and Eq.~(\ref{eq:6}), which is, however, not observed in our DCA results. This absence of significant effects of $d-$wave fluctuations in the self-energy can be understood from a geometrical perspective: In fact, to get large $d-$wave pairing fluctuations, the sign of the expectation value in Eq.~(\ref{eq:6})  must vary with ${\bf K}'$ in accordance to $f({\bf K}')$. Then, however, for a purely local interaction, the contribution to Eq.~(\ref{eq:5}) cannot be large, since the factor $f({\bf K}')$ is not present there. Hence, we do not expect superconductivity fluctuations to contribute efficiently to $\Sigma$ in general, except for $s-$wave superconductivity (i.e., $f({\bf K}) = 1$), as it is indeed the case of the $U\!<\!0$ Hubbard model.